\documentclass[twocolumn,showpacs,preprintnumbers,amsmath,amssymb]{revtex4}

\usepackage{graphicx}
\usepackage{dcolumn}
\usepackage{bm}

\begin{document}

\title{Equilibrium long-ranged charge correlations at the surface \\
of a conductor coupled to the electromagnetic radiation}
\author{Ladislav \v{S}amaj}
\altaffiliation[On leave from ]
{Institute of Physics, Slovak Academy of Sciences,
Bratislava}
\email{Ladislav.Samaj@savba.sk}
\author{Bernard Jancovici}
\email{Bernard.Jancovici@th.u-psud.fr}
\affiliation{Laboratoire de Physique Th\'eorique,
Universit\'e de Paris-Sud \\
91405 Orsay Cedex, France\footnote{Unit\'e Mixte de Recherche No 8627-CNRS}}

\date{\today}

\begin{abstract}
This study is related to the fluctuation theory of electromagnetic
fields, charges and currents.
The three-dimensional system under consideration is 
a semi-infinite conductor, modeled by the jellium, in vacuum.
In previous theoretical studies it was found that the correlation functions 
of the surface charge density on the conductor decay as the inverse cube 
of the distance at asymptotically large distances.
The prefactor to this asymptotic decay was obtained in the classical
limit and in the quantum case without retardation effects.
To describe the retarded regime, we study a more general problem
of the semi-infinite jellium in thermal equilibrium with the radiated
electromagnetic field.
By using Rytov's fluctuational electrodynamics we show that, for both 
static and time-dependent surface charge correlation functions, 
the inclusion of retardation effects causes the quantum prefactor 
to take its universal static classical form, for any temperature.
\end{abstract}


\maketitle

\section{Introduction}
Experimental and theoretical investigations of charged systems in
thermal equilibrium play a key role in the understanding of various 
fundamental properties of solids and liquids, in the bulk as well as
on the surface.
There exist two complementary theoretical approaches to Coulomb models:
one based on the solution of microscopic models, the other
based on the assumption of validity of macroscopic electrodynamics.
The microscopic description is more laborious and complicated but,
if available, it can reveal restricted applicability of 
the macroscopic theory.
On the other hand, the macroscopic phenomenology usually allows us 
to predict, with much less effort, basic features of relatively 
complicated complex physical systems. 

In this paper, we study the fluctuations of electromagnetic fields,
charges and currents in systems formulated in the three-dimensional (3D) 
Cartesian space of points ${\bf r}=(x,y,z)$.
We shall deal with semi-infinite geometries, inhomogeneous say
along the first coordinate $x$. 
It will be sometimes useful to denote the remaining two coordinates normal 
to $x$ as ${\bf R}=(y,z)$.
The physical situation we are interested in is pictured in Fig.1.
In its most general formulation, the model consists of two semi-infinite
media with the frequency-dependent dielectric functions 
$\epsilon_1(\omega)$ and $\epsilon_2(\omega)$, in the half-space 
$\Lambda_1=\{ {\bf r}=(x,{\bf R}); x>0\}$ and in the complementary 
half-space $\Lambda_2=\{ {\bf r}=(x,{\bf R}); x<0\}$, respectively.
The interface between the media is localized at $x=0$.
Although we shall derive basic formulas for this general case,
the analysis of the results will be done for the specific case
when the half-space $\Lambda_1$ is occupied by a Coulomb fluid,
namely a jellium, and the half-space $\Lambda_2$ is formed by
an impenetrable plain hard wall with the vacuum permittivity. 
For a Coulomb fluid composed of charged particles, the average 
particle density is varying at microscopic distances 
(of the order of the correlation length) from the interface.
Since we shall be interested in macroscopic electrodynamic phenomena 
at distances much larger than the microscopic length scale, 
we can assume that the Coulomb fluid and the dielectric function
are homogeneous in the whole half-space $\Lambda_1$.

\begin{figure}[t]
\begin{center}
\includegraphics[width=0.45\textwidth,clip]{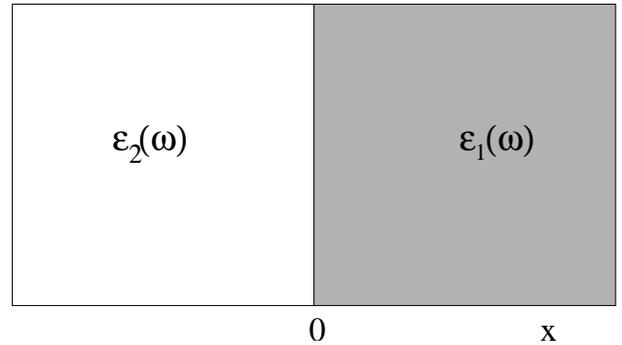}
\caption{Two semi-infinite media characterized by dielectric functions
$\epsilon_1(\omega)$ and $\epsilon_2(\omega)$.}
\end{center}
\end{figure}

The presence of the Coulomb fluid gives rise to a surface charge density 
$\sigma$ on the conductor which has to be understood as being the
microscopic volume charge density integrated on some microscopic depth.
It is associated with the discontinuity of the electric field ${\bf E}$ 
at the surface $x=0$ of the conductor.
At a point ${\bf r}=(0,{\bf R})$ on the surface, in Gauss units,
\begin{equation} \label{1}
4 \pi \sigma({\bf R}) = E_x^{+}({\bf R}) - E_x^{-}({\bf R}) ,
\end{equation} 
where the upperscript $+$ $(-)$ means approaching the surface through 
the limit $x\to 0^+$ ($x\to 0^-$).
The quantity of interest is the surface charge density correlation function
between two points 
\begin{eqnarray} 
& & \langle \sigma({\bf R}) \sigma({\bf R}') \rangle^{\rm T} = 
\frac{1}{(4\pi)^2} \nonumber \\ & & \qquad \times
\left\langle \left[ E_x^{+}({\bf R}) - E_x^{-}({\bf R}) \right]
\left[ E_x^{+}({\bf R}') - E_x^{-}({\bf R}') \right] \right\rangle^{\rm T} , 
\quad \label{2}
\end{eqnarray}
where $\langle \cdots \rangle^{\rm T}$ represents a truncated equilibrium
statistical average, $\langle A B \rangle^{\rm T} = \langle A B \rangle 
- \langle A \rangle \langle B \rangle$, at the inverse temperature $\beta$.
The system is translationally invariant in the ${\bf R}$-plane and
so the charge correlation function depends only on the distance 
between the points,
\begin{equation} \label{3}
\langle \sigma({\bf R}) \sigma({\bf R}') \rangle^{\rm T} =
S(\vert {\bf R}-{\bf R}'\vert) .
\end{equation}
It is useful to introduce the Fourier transform
\begin{equation} \label{4}
S({\bf q}) = \int d^2 R\, e^{i{\bf q}\cdot {\bf R}}
S({\bf R}) ,
\end{equation}
with ${\bf q}=(q_y,q_z)$ being the 2D wave vector.

For classical Coulomb fluids composed of charged particles with
the instantaneous Coulomb interactions, $S(R)$ can be retrieved
by simple macroscopic argument based on the electrostatic method 
of images, giving \cite{Jancovici95}
\begin{equation} \label{5}
\beta S_{\rm cl}(R) = - \frac{1}{8 \pi^2} \frac{1}{R^3} ,
\end{equation}
where the distance $R = \vert {\bf R} \vert$ is much larger than
any microscopic length scale, like the particle correlation length.
Since, in the sense of distributions, the 2D Fourier transform of
$1/R^3$ is $-2\pi q$, the function $\beta S_{\rm cl}(q)$ has 
a kink singularity at $q=0$,
\begin{equation} \label{6}
\beta S_{\rm cl}(q) \sim \frac{1}{4\pi} q , \qquad q\to 0 .
\end{equation}
Note the universal form of the asymptotic formulas, independent of 
the composition of the Coulomb fluid, which is related to specific 
sum rules for systems with Coulomb interaction 
(for a review, see Ref. \onlinecite{Martin88}).
We see that the surface charge correlation on the conductor
has a long-ranged decay.
Such behavior is in contrast with a short-ranged, usually exponential,
decay which occurs in the bulk charge density correlations.
The same result (\ref{5}) has been obtained also 
in the microscopic language \cite{Jancovici82}.
The macroscopic formulas for the surface charge density correlation functions 
on conductors of various shapes were derived by 
Choquard et al \cite{Choquard89}.
The inverse-power law behavior of the surface charge density of type
(\ref{5}) causes the conductor-shape dependence of the dielectric
susceptibility tensor (which relates the average polarization of the system
to a constant applied electric field, in the linear limit)
\cite{Choquard89,Jancovici04}.
Such a phenomenon is predicted by the macroscopic laws of electrostatics
\cite{Landau60}.

The extension of the classical result (\ref{5}), or equivalently (\ref{6}),
to a quantum Coulomb fluid was accomplished in Ref. \onlinecite{Jancovici85a}.
The model under consideration was the jellium,
i.e. a system of pointlike particles of charge $e$, mass $m$ and
bulk number density $n$, immersed in a uniform neutralizing background
of charge density $- e n$; the background is assumed to have
the vacuum dielectric constant 1.
The dynamical properties of the jellium have a special feature:
there is no viscous damping of the long-wavelength plasma oscillations
for identically charged particles.
The absence of damping was crucial in the derivation of the large-distance
behavior of the surface charge correlation function using long-wavelength
collective modes, namely the nondispersive (if the pressure term is neglected)
bulk plasmons with frequency $\omega_p$ and the surface plasmons with 
frequency $\omega_s$, given by
\begin{equation} \label{7}
\omega_p = \left( \frac{4\pi n e^2}{m} \right)^{1/2}, \qquad
\omega_s = \frac{\omega_p}{\sqrt{2}} .
\end{equation}
In the Fourier space, the obtained asymptotic $q\to 0$ result 
has the non-universal form \cite{Jancovici85a}
\begin{subequations}
\begin{eqnarray} 
\beta S_{\rm qu}(q) & \sim & \frac{[2 f(\omega_s) - f(\omega_p)]}{4\pi}\, 
q , \qquad q\to 0 , \label{8a} \\
f(\omega) & = & \frac{\beta\hbar\omega}{2}
\coth\left( \frac{\beta\hbar\omega}{2} \right) . \label{8b}
\end{eqnarray}
\end{subequations}
According to the correspondence principle, a quantum system admits
the classical treatment in the high-temperature region, which
corresponds in our case to $\beta\hbar\to 0$.
In this limit, the function $f(\omega)=1$ for any $\omega$
and the quantum relation (\ref{8a}) indeed reduces 
to the classical one (\ref{6}). 

The surface charge correlation function discussed up to now was static,
i.e. time taken at the two distinct points was the same.
The generalization to the time-dependent correlation function is
accomplished by introducing a time-dependent surface charge density,
defined as the Heisenberg operator
\begin{equation} \label{9}
\sigma(t,{\bf R}) = \exp(i H t/\hbar) \sigma({\bf R}) \exp(-i H t/\hbar)
\end{equation}
with $H$ being the Hamiltonian of the Coulomb system.
Since the Heisenberg operators at different times do not commute, 
it is useful to introduce the time-dependent
surface charge correlation function in a symmetrized form
\begin{equation} \label{10}
[ \sigma({\bf R}) \sigma({\bf R}') ]_t \equiv \frac{1}{2}
\left\langle \sigma(t,{\bf R}) \sigma(0,{\bf R}') +
\sigma(0,{\bf R}') \sigma(t,{\bf R}) \right\rangle^{\rm T} . 
\end{equation}  
This symmetrized correlation function still possesses the translational
invariance in the ${\bf R}$-plane, 
\begin{equation} \label{11}
[ \sigma({\bf R}) \sigma({\bf R}') ]_t = 
S_{\rm qu}(t,\vert {\bf R}-{\bf R}'\vert ) .
\end{equation}
The asymptotic $q\to 0$ formula was shown 
\cite{Jancovici85a,Jancovici85b} to be 
\begin{equation} \label{12}
\beta S_{\rm qu}(t,q) \mathop{\sim}_{q\to 0} 
\frac{[2 f(\omega_s)\cos(\omega_s t) 
- f(\omega_p)\cos(\omega_p t)]}{4\pi}\, q . 
\end{equation}
For $t=0$, this formula reduces to the static one (\ref{8a}).
Note that in the classical limit with $f(\omega)=1$ for any $\omega$,
the surface charge correlation function is non-universal and
exhibits a periodic time dependence of type
\begin{equation} \label{13}
\beta S_{\rm cl}(t,q) \mathop{\sim}_{q\to 0} 
\frac{[2 \cos(\omega_s t)-\cos(\omega_p t)]}{4\pi}\, q .
\end{equation}

The quantum static and time-dependent results (\ref{8a}) and 
(\ref{12}), respectively, were derived in the {\em nonretarded} regime
where the speed of light $c$ is taken to be infinitely large, $c=\infty$.
The effects of the retardation take place just at large distances
we are interested in, and so they should be taken into account.
To describe the {\em retarded} regime with the finite value of $c$,
we study a more general problem of the semi-infinite Coulomb system 
in thermal equilibrium with the radiated electromagnetic (EM) field.

We intend to treat both the Coulomb fluid and the radiation like
quantum objects.
For the static case, a substantial simplification arises in 
the high-temperature limit $\beta\hbar\to 0$.
First, according to the correspondence principle, both matter and
radiation can be treated classically.
Second, the application of the Bohr-van Leeuwen theorem
\cite{Bohr11,VanLeeuwen21} leads to the decoupling between classical
matter and radiation, and to an effective elimination of the magnetic
forces in the matter; for a detailed treatment of this subject, see 
Ref. \onlinecite{Alastuey00}.
This means that the matter can be treated as a classical matter, 
unaffected by radiation, where the charges interact only via 
the instantaneous Coulomb potential.
Because of the absence of quantum effects for $\beta\hbar\to 0$ we expect 
that the classical formula (\ref{6}) will be restored in this limit.
How the retardation effects manifest themselves at a finite temperature 
is an open question which is studied in this paper.
All that has been said in this paragraph does not apply, in general,
to time-dependent quantities \cite{Alastuey08}.
In particular, in the presence of the radiation, it can happen that
the classical time-dependent result (\ref{13}) is not reproduced
in the limit $\beta\hbar\to 0$. 

To deal with the proposed physical problem of such complexity,  
we shall use a macroscopic theory of equilibrium thermal fluctuations
of EM field, published by Rytov \cite{Rytov58} and further developed
in Ref. \onlinecite{Levin67}. 
This theory is presented also in the Course of Theoretical Physics 
by Landau and Lifshitz \cite{Lifshitz80} of which we adopt the notation.
Although the EM fluctuation theory was extensively applied in
investigations of thermally excited surface EM waves 
\cite{Joulain05,Bimonte06}, we did not find a study about 
the present topic concerning the long-range decay of surface
charge correlation functions. 

The main results obtained in this paper are the following.
The quantum static and time-dependent prefactors of the small-$q$ 
behaviors (\ref{8a}) and (\ref{12}), respectively, remain valid 
at some intermediate distances, 
where the retardation effects do not play any role.
In the strict large-distance asymptotic limit, for both static and 
time-dependent surface charge correlation functions, the inclusion of
the retardation effects causes the quantum prefactor to take its
static classical form, see Eqs. (\ref{5}) or (\ref{6}).

The paper is outlined as follows.
In Sec. II, we review shortly the EM fluctuation theory and derive 
the formula for the surface charge correlation function between
two semi-infinite media.
The analysis of the asymptotic form of this formula for the
configuration of interest jellium-vacuum is the subject of Sec. III.
A brief recapitulation and concluding remarks are given in Sec. IV.

\section{Derivation of basic formulae}
We consider the $(3+1)$-dimensional space with spatial vectors 
${\bf r}$ and time $t$.
The physical system of interest is a medium and an EM field present in it, 
which are in thermodynamic equilibrium.

The medium is composed of moving charged particles which
are assumed to be non-relativistic.
In the long-wavelength scale much larger that the interparticle
distances in the medium, its isotropic macroscopic characteristics are 
the frequency and (possibly) position dependent dielectric function 
$\epsilon(\omega;{\bf r})$ and permeability $\mu(\omega;{\bf r})$.
We shall assume that the medium has no magnetic structure, i.e.
it is not magnetoactive, and $\mu=1$.

The matter is coupled to the EM field. 
The {\em classical} EM field potentials form a 4-vector $(\phi,{\bf A})$, 
where $\phi(t,{\bf r})$ is the scalar potential and ${\bf A}(t,{\bf r})$ 
is the vector potential.
In the considered Weyl gauge $\phi = 0$, the microscopic electric 
and magnetic fields are given by
\begin{equation} \label{14}
{\bf E} = - \frac{1}{c} \frac{\partial {\bf A}}{\partial t} , \qquad
{\bf B} = {\rm curl}\, {\bf A} .
\end{equation}
The elementary excitations of the {\em quantized} EM field are described 
by the photon operators $\hat{A}_j$ $(j=x,y,z)$ which are self-conjugate 
Bose operators.

Medium and coupled radiation are in thermal equilibrium.
The EM fields and inductions are random variables which
fluctuate around their mean values. 
These mean values are the quantities obeying macroscopic Maxwell's equations.
The construction of all types of photon Green's functions is based on 
the retarded Green function, defined as the tensor
\begin{widetext}
\begin{equation} 
i D_{jk}(t;{\bf r},{\bf r}') = \left\{
\begin{array}{lr} 
\langle \hat{A}_j(t,{\bf r}) \hat{A}_k(0,{\bf r}') -
\hat{A}_k(0,{\bf r}') \hat{A}_j(t,{\bf r}) \rangle , & t>0 ,\cr
& \cr 0 , & t<0 .
\end{array} \right. \label{15}
\end{equation}
\end{widetext}
Here, $j,k=x,y,z$, $\hat{A}_j(t,{\bf r})$ denotes the vector-potential 
operator in the Heisenberg picture and the angular brackets represent 
equilibrium averaging over the Gibbs distribution of the whole system.
In what follows, we shall work in the Fourier space with respect to time.
The Fourier transform of the retarded Green function reads
\begin{equation} \label{16}
D_{jk}(\omega;{\bf r},{\bf r}') = \int_0^{\infty} d t\,
e^{i\omega t} D_{jk}(t;{\bf r},{\bf r}') .
\end{equation}
For media with no magnetic structure, the Green function tensor possesses
the symmetry
\begin{equation} \label{17}
D_{jk}(\omega;{\bf r},{\bf r}') = 
D_{kj}(\omega;{\bf r}',{\bf r}) 
\end{equation}

Within the framework of the fluctuational electrodynamics of Rytov
\cite{Rytov58,Levin67,Lifshitz80}, the retarded Green function tensor
fulfills the differential equation
\begin{eqnarray}
\sum_{l=1}^3 \left[ \frac{\partial^2}{\partial x_j \partial x_l}
- \delta_{jl} \Delta - \delta_{jl} \frac{\omega^2}{c^2} 
\epsilon(\omega;{\bf r}) \right] D_{lk}(\omega;{\bf r},{\bf r}') 
\nonumber \\
= - 4 \pi \hbar \delta_{jk} \delta({\bf r}-{\bf r}') . \label{18}
\end{eqnarray} 
Here, in order to simplify the notation, the vector ${\bf r}=(x,y,z)$ 
is represented as $(x_1,x_2,x_3)$.
The differential equation (\ref{18}) must be supplemented by 
certain boundary conditions.
The second space variable ${\bf r}'$ and the second index $k$ are not
involved in the mathematical operations on the tensor, so they only
act as parameters.
The boundary conditions are thus formulated with respect to the
coordinate ${\bf r}$ and the Green function 
$D_{lk}(\omega;{\bf r},{\bf r}')$ is considered as a vector
with the components $l=x,y,z$.
There is an obvious boundary condition of regularity at infinity,
$\vert {\bf r}\vert \to \infty$.
At an interface between two different media, the boundary conditions
correspond to the macroscopic requirements that the tangential
components of the fields ${\bf E}$ and ${\bf H}={\bf B}$ be continuous.
Since the electric and magnetic fields are related to the vector potential 
by (\ref{14}), the role of the vector components $E_l$, up to an irrelevant
multiplicative constant, is played by the quantity
\begin{equation} \label{19}
i \frac{\omega}{c} D_{lk}(\omega;{\bf r},{\bf r}')
\end{equation}
and the role of the vector component $H_l$ is played by the quantity
\begin{equation} \label{20}
\sum_j {\rm curl}_{lj} D_{jk}(\omega;{\bf r},{\bf r}') .
\end{equation}
Here, we use the notation
${\rm curl}_{lj} = \sum_m e_{lmj}\partial/\partial x_m$ with
$e_{lmj}$ being the unit antisymmetric pseudo-tensor. 
In both cases (\ref{19}) and (\ref{20}), the tangential components,
which are continuous at the interface, correspond to indices $l=y,z$.

The fluctuation-dissipation theorem tells us that the fluctuations
of random variables can be expressed in terms of the corresponding
susceptibilities.
For the assumed symmetry (\ref{17}), the theorem implies
\begin{equation} \label{21}
[ E_j({\bf r}) E_k({\bf r}') ]_{\omega} =    
- \frac{\omega^2}{c^2} \coth( \beta\hbar\omega/2 )\,  
{\rm Im}\, D_{jk}(\omega;{\bf r},{\bf r}') , 
\end{equation}
where the spectral distribution of the electric-field fluctuations
$[ E_j({\bf r}) E_k({\bf r}') ]_{\omega}$ is the Fourier transform in time 
of the symmetrized (truncated) correlation function
\begin{equation} \label{22}
\frac{1}{2} \left\langle \hat{E}_j(t,{\bf r}) \hat{E}_k(0,{\bf r}')
+ \hat{E}_k(0,{\bf r}') \hat{E}_j(t,{\bf r}) \right\rangle^{\rm T} .
\end{equation}

Since the studied problem of two semi-infinite media, presented in Fig. 1,
is translationally invariant in the ${\bf R}$-plane perpendicular to 
the $x$ axis, we introduce the Fourier transform of the Green function
tensor with the wave vector ${\bf q}=(q_y,q_z)$:
\begin{equation} \label{23}
D_{jk}(\omega;{\bf r},{\bf r}') = \int \frac{d^2q}{(2\pi)^2}
e^{- i {\bf q}\cdot({\bf R}-{\bf R}')}
D_{jk}(\omega,{\bf q};x,x') .
\end{equation} 
At the time being, the dielectric functions $\epsilon_1(\omega)$ and 
$\epsilon_2(\omega)$ are taken as general. 
The Green function tensor for simple planar systems was obtained, as 
the solution of the differential equation (\ref{18}) supplemented by 
the mentioned boundary conditions, in a number of papers, 
see, e.g., Appendix A of Ref. \cite{Joulain05} or, for multilayers, 
Ref. \cite{Tomas95}.
We shall not repeat the derivation, but only write the final formulae.
Let us define for each of the half-space regions the inverse length 
$\kappa_j(\omega,q)$ $(j=1,2)$ by
\begin{equation} \label{24}
\kappa_j^2(\omega,q) = q^2 - \frac{\omega^2}{c^2} \epsilon_j(\omega) ,
\qquad {\rm Re}\, \kappa_j(\omega,q) > 0 ;
\end{equation}
from two possible solutions for $\kappa_j$ we choose the one with the 
positive real part in order to ensure the regularity of the Green functions 
at asymptotically large distances from the interface (see below).
We shall only need the quantity $D_{xx}(\omega,q;x,x')$ for which the 
previously obtained results can be summarized as follows:
\begin{itemize}
\item
If $x,x'>0$,
\begin{eqnarray}
D_{xx} & = & \frac{4\pi\hbar c^2}{\omega^2\epsilon_1} \delta(x-x')
- \frac{2\pi\hbar(cq)^2}{\omega^2\epsilon_1 \kappa_1} \nonumber \\
& & \times \left[ e^{-\kappa_1\vert x-x'\vert} +
\frac{\kappa_1\epsilon_2-\kappa_2\epsilon_1}{\kappa_1\epsilon_2
+\kappa_2\epsilon_1} e^{-\kappa_1(x+x')} \right] . \label{25}
\end{eqnarray}
\item
If $x<0$ and $x'>0$,
\begin{equation} \label{26}
D_{xx} = - \frac{4\pi\hbar (cq)^2}{\omega^2}
\frac{1}{\kappa_1\epsilon_2+\kappa_2\epsilon_1} e^{\kappa_2 x - \kappa_1 x'} . 
\end{equation}
\item
The case $x>0$ and $x'<0$ is deducible from Eq. (\ref{26}) by using
the symmetry relation (\ref{17}).
\item
If $x,x'<0$, considering the $1\leftrightarrow 2$ media exchange symmetry,
we obtain from Eq. (\ref{25}) that
\begin{eqnarray}
D_{xx} & = & \frac{4\pi\hbar c^2}{\omega^2\epsilon_2} \delta(x-x')
- \frac{2\pi\hbar(cq)^2}{\omega^2\epsilon_2 \kappa_2} \nonumber \\
& & \times \left[ e^{-\kappa_2\vert x-x'\vert} +
\frac{\kappa_2\epsilon_1-\kappa_1\epsilon_2}{\kappa_2\epsilon_1
+\kappa_1\epsilon_2} e^{\kappa_2(x+x')} \right] . \label{27}
\end{eqnarray}
\end{itemize}

The symmetrized surface charge correlation function (\ref{10})
is expressible in terms of the symmetrized $xx$ electric-field
fluctuations by using an obvious analogy of relation (\ref{2}).
These electric-field fluctuations are related to the $xx$ elements
of the Green function tensor via Eq. (\ref{21}).
The terms proportional to $\delta(x-x')$ in Eqs. (\ref{25}) 
and (\ref{27}) can be ignored since they originate from
the short-distance terms proportional to $\delta({\bf r}-{\bf r}')$
which do not play any role in the large-distance asymptotic.
Since the combination
\begin{eqnarray}
D_{xx}(0^+,0^+) + D_{xx}(0^-,0^-) - 2 D_{xx}(0^+,0^-) \nonumber \\
= - \frac{4\pi\hbar (cq)^2}{\omega^2} 
\frac{1}{\kappa_1\epsilon_2+\kappa_2\epsilon_1}
\left( \frac{\epsilon_2}{\epsilon_1} + \frac{\epsilon_1}{\epsilon_2}
- 2 \right) , \label{28}
\end{eqnarray}
we finally get
\begin{subequations} \label{29}
\begin{eqnarray} 
\beta S(\omega,q) & = & \frac{\beta\hbar}{4\pi} \coth(\beta\hbar\omega/2)
q^2 {\rm Im}\, g(\omega,q) , \label{29a} \\
g(\omega,q) & = & 
\frac{1}{\kappa_1(\omega,q)\epsilon_2(\omega)
+\kappa_2(\omega,q)\epsilon_1(\omega)} \nonumber \\ & & \times
\frac{[\epsilon_1(\omega)-\epsilon_2(\omega)]^2}{\epsilon_1(\omega)
\epsilon_2(\omega)} . \label{29b}
\end{eqnarray}
\end{subequations}
Our task is to determine, for given $\epsilon_1(\omega)$ and
$\epsilon_2(\omega)$, the small-$q$ behavior of the function
\begin{equation} \label{30}
\beta S(t,q) = \int_{-\infty}^{\infty} \frac{d\omega}{2\pi}
e^{-i\omega t} \beta S(\omega,q) .
\end{equation}

\section{Analysis of asymptotic behavior}
In any material medium, the complex dielectric function $\epsilon(\omega)$
possesses the symmetry \cite{Jackson75}
\begin{equation} \label{31}
\epsilon^*(\omega) = \epsilon(-\omega) , \qquad \omega\in R .
\end{equation}
Denoting $\epsilon(\omega) = \epsilon'(\omega) + i \epsilon''(\omega)$,
where both the real $\epsilon'(\omega)$ and imaginary $\epsilon''(\omega)$ 
parts are real numbers, this means that 
\begin{equation} \label{32}
\epsilon'(\omega) = \epsilon'(-\omega) , \qquad
\epsilon''(\omega) = -\epsilon''(-\omega) .
\end{equation}
The dielectric function has many other general properties.
Like for instance, for any material medium with absorption it holds
\begin{equation} \label{33}
\epsilon''(\omega) > 0 \qquad \mbox{when $\omega>0$.} 
\end{equation}

We shall analyze the general fluctuation results of the previous
section for the model configuration of interest: the semi-infinite
jellium in vacuum.
As concerns the jellium, the dissipation goes to zero in 
the limit of small wave numbers \cite{Pines66}.
The dielectric function can be thus shown to be \cite{Pitarke07} 
the Drude one
\begin{equation} \label{34}
\epsilon_1(\omega) = 1 - \frac{\omega_p^2}{\omega(\omega + i\eta)} ,
\end{equation}
where $\omega_p$ is the plasma frequency defined in Eq. (\ref{7}) and
the dissipation constant $\eta$ is taken as positive infinitesimal,
$\eta\to 0^+$.  
The Weierstrass theorem reads
\begin{equation} \label{35}
\lim_{\eta\to 0^+} \frac{1}{x\pm i \eta} = 
{\cal P}\left( \frac{1}{x}\right) \mp i\pi\delta(x) ,\qquad
x\in R ,
\end{equation} 
where ${\cal P}$ denotes the Cauchy principal value.
It is tempting to apply this theorem directly to the representation 
(\ref{34}), with the result
\begin{equation} \label{36}
\epsilon'_1(\omega) = 1 - \omega_p^2 {\cal P} 
\left( \frac{1}{\omega^2} \right) , \qquad
\epsilon''_1(\omega) = \pi \omega_p^2 \frac{1}{\omega} \delta(\omega) .
\end{equation}
Although both real and imaginary parts satisfy the necessary conditions
(\ref{32}) and (\ref{33}), the expression for the imaginary part
has no meaning.
In the algebraic manipulations with $\epsilon_1(\omega)$, we must therefore
keep the positive infinitesimal $\eta$ in the representation (\ref{34}) 
up to the end and to apply the Weierstrass theorem (\ref{35}) only
to the final formula.
In the vacuum region, 
\begin{equation} \label{37}
\epsilon_2(\omega) = 1 .
\end{equation}
To determine the sign of some quantities, we shall sometime need 
the infinitesimal imaginary part of the vacuum dielectric constant.
To fulfill the required properties (\ref{32}) and (\ref{33}),
the vacuum dielectric constant in fact corresponds to the limit
\begin{equation} \label{38}
\epsilon_2(\omega) = 1 + i\, {\rm sgn}(\omega) 0^+ .
\end{equation}

It is evident that the inverse length $\kappa$, defined by
the relation (\ref{24}), also possesses the symmetry
\begin{equation} \label{39}
\kappa^*(\omega,q) = \kappa(-\omega,q) , \qquad \omega\in R .
\end{equation} 
In terms of the real and imaginary parts, 
$\kappa(\omega,q) = \kappa'(\omega,q) + i \kappa''(\omega,q)$, 
this symmetry is equivalent to
\begin{equation} \label{40}
\kappa'(\omega,q) = \kappa'(-\omega,q) , \qquad
\kappa''(\omega,q) = - \kappa''(-\omega,q) .
\end{equation}

\subsection{Nonretarded limit}
In the nonretarded limit $c=\infty$, the definition (\ref{24})
becomes $\kappa_1=\kappa_2=q$ and Eq. (\ref{29b}) reduces to
\begin{equation} \label{41}
g(\omega,q) = \frac{1}{q} \left[
\frac{1}{\epsilon_1(\omega)} + \frac{1}{\epsilon_2(\omega)} -
\frac{4}{\epsilon_1(\omega)+\epsilon_2(\omega)} \right] .
\end{equation}
Substituting here the dielectric functions (\ref{34}) and (\ref{37}), 
and then applying the Weierstrass theorem (\ref{35}), we obtain
\begin{eqnarray}
{\rm Im}\, g(\omega,q) & = & \frac{\pi\omega_p^2}{q} {\rm sgn}(\omega)
\left[ \delta(\omega^2-\omega_s^2) - \delta(\omega^2-\omega_p^2) \right]
\nonumber \\ & = & \frac{\pi}{q} \omega
\left[ 2 \omega_s \delta(\omega^2-\omega_s^2) - 
\omega_p \delta(\omega^2-\omega_p^2) \right] ; \qquad \label{42}
\end{eqnarray}
the frequency of surface plasmons $\omega_s$ is defined in Eq. (\ref{7}).
Using the general formula for the $\delta$-functions
\begin{equation} \label{43}
\delta[h(x)] = \sum_j \frac{\delta(x-x_j)}{\vert h'(x_j)\vert} ,
\end{equation}
where $\{ x_j\}$ are the real roots of $h(x)$, it is a simple task
to show that relations (\ref{29a}) and (\ref{30}) imply
the expected result (\ref{12}).

The $c=\infty$ treatment of our equations is mathematically accessible
and reproduces the previous ``nonretarded'' results.
Since the true value of $c$ is finite, there exists a limitation for
wave numbers $q$ or distances $\lambda\sim 1/q$ for which retardation 
effects are negligible.
We can find this limitation via a simple dimensional analysis
of Eqs. (\ref{29}) and (\ref{30}).
The invoked substitution $\beta\hbar\omega=\omega'$ tells us
that there are two dimensionless quantities in the theory:
\begin{equation} \label{44}
u = \beta \hbar c q , \qquad v = \beta \hbar \omega_p .
\end{equation}
The requirement of the smallness of $u\ll 1$ is equivalent to
the natural condition
\begin{equation} \label{45}
\lambda_{\rm ph} \ll \lambda ,
\end{equation}
where $\lambda_{\rm ph} \propto \beta\hbar c$ is the thermal
de Broglie wavelength of photon.
The nonretarded limit $c\to\infty$ describes adequately the region
corresponding to the inequality $u\gg v$, i.e.
\begin{equation} \label{46}
\lambda \ll \frac{c}{\omega_p} .
\end{equation}
For rough value of the free electron density in metal
$n\sim 10^{29} {\rm m}^{-3}$, taking $e$ and $m$ as the charge
and mass of electron, we have $\omega_p\sim 10^{15} {\rm s}^{-1}$ and so
\begin{equation} \label{47}
\lambda \ll 10^{-7} {\rm m} \sim 10^2 d ,
\end{equation}
where $d\sim 10^{-9} {\rm m}$ is the mean interparticle distance.
We see that the distance, over which the retardation effects 
are negligible and so the formula (\ref{12}) is adequate, 
is relatively large.

\subsection{Retarded region}
If we want to deal strictly with the large-distance
asymptotic behavior of the surface charge correlation function,
we must take $c$ as a finite number and consider the region
corresponding to the inequality $u\ll v$, or $c q \ll \omega_p$.
This requires a detailed analysis of Eqs. (\ref{29}) and (\ref{30}).

We start with an algebraic treatment of the function $g(\omega,q)$
defined by Eq. (\ref{29b}).
Multiplying both numerator and denominator by 
$\kappa_1\epsilon_2 - \kappa_2\epsilon_1$, and using the equality
\begin{equation} \label{48}
\kappa_1^2 \epsilon_2^2 - \kappa_2^2 \epsilon_1^2 = (\epsilon_2-\epsilon_1)
\left[ q^2 (\epsilon_1+\epsilon_2) - 
\frac{\omega^2}{c^2} \epsilon_1 \epsilon_2 \right] ,
\end{equation}
we obtain
\begin{equation} \label{49}
g = \frac{\kappa_1\epsilon_2 - \kappa_2\epsilon_1}{q^2 (\epsilon_1+\epsilon_2) 
- (\omega^2/c^2) \epsilon_1 \epsilon_2}
\left( \frac{1}{\epsilon_1} - \frac{1}{\epsilon_2} \right) . 
\end{equation}
The denominator of this expression for $g$ is related to the
surface-plasmon dispersion relation
\begin{equation} \label{50}
q^2 (\epsilon_1+\epsilon_2) - \frac{\omega^2}{c^2} 
\epsilon_1 \epsilon_2 = 0 ;
\end{equation}
for a review, see Ref. \onlinecite{Pitarke07}.
Considering the metal dielectric function $\epsilon_1(\omega)$, 
as defined by Eq. (\ref{34}) but for the time being with $\eta=0$, 
and the vacuum one $\epsilon_2(\omega)=1$, Eq. (\ref{50}) has two solutions  
\begin{equation} \label{51}
\omega_{\pm}^2(q) = (\omega_p^2/2) + (cq)^2 \pm
\left[ \left( \omega_p^2/2 \right)^2 + (cq)^4 \right]^{1/2} .
\end{equation}
These two dispersion relations are represented, for $\omega\ge 0$, 
in Fig. 2 by solid lines, together with the (dashed) light line $\omega=c q$.
The upper solid line $\omega_+$ always lies above the dispersion curve 
of light in the metal \cite{Jackson75}, 
\begin{figure}[b]
\begin{center}
\includegraphics[width=0.45\textwidth,clip]{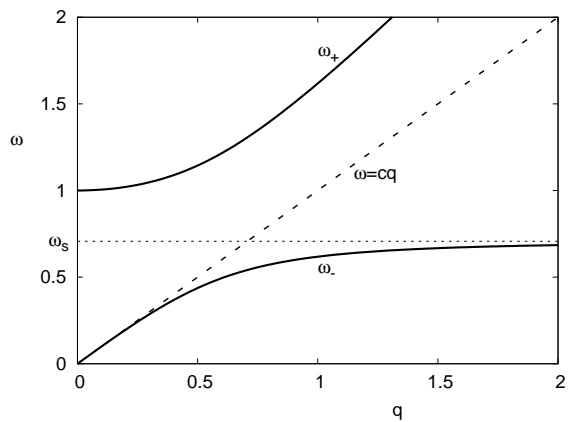}
\caption{Dispersion (solid) curves $\omega_-(q)$, $\omega_+(q)$ and the
light (dashed) line $\omega=cq$; the plane vector $q$ is represented 
in units of $\omega_p/c$, the frequency $\omega$ in units of $\omega_p$.}
\end{center}
\end{figure}
\begin{equation} \label{52}
\omega_+^2(q)>\omega_p^2+(cq)^2.
\end{equation}
The lower solid line $\omega_-(q)$, corresponding to the dispersion relation 
of the surface-plasmon polariton, lies always below the light line,
\begin{equation} \label{53}
\omega_-^2(q)< (cq)^2.
\end{equation}
In the nonretarded limit $cq\gg \omega_s$, it approaches the
nondispersive surface-plasmon frequency $\omega_s=\omega_p/\sqrt{2}$.
In the retarded region of interest $cq\ll \omega_s$, it approaches
the light line $\omega = c q$.
In what follows, we shall need the small-$q$ expansions
\begin{subequations} \label{54}
\begin{eqnarray}
\omega_-^2(q) & = & (cq)^2 - \frac{(cq)^4}{\omega_p^2} - \cdots , 
\label{54a} \\
\omega_+^2(q) & = & \omega_p^2 + (cq)^2 + \frac{(cq)^4}{\omega_p^2} 
+ \cdots . \label{54b}
\end{eqnarray}
\end{subequations}

After substituting the metal dielectric function (\ref{34}), now
completely with $\eta$ being positive infinitesimal, and 
the vacuum dielectric function $\epsilon_2(\omega)=1$ into (\ref{49}), 
we obtain
\begin{equation} \label{55}
g(\omega,q) = g_1(\omega,q) + g_2(\omega,q) ,
\end{equation}
where
\begin{subequations} \label{56}
\begin{eqnarray}
g_1 & = & - \frac{(c\omega_p)^2}{\omega^2-\omega_p^2+ i \omega\eta}
\nonumber \\ & & \times
\frac{\kappa_1 \omega (\omega+ i \eta)}{(\omega^2-\omega_-^2)
(\omega^2-\omega_+^2) + i \omega\nu\eta} , \label{56a} \\
g_2 & = & (c\omega_p)^2 \frac{\kappa_2}{(\omega^2-\omega_-^2)
(\omega^2-\omega_+^2) + i \omega\nu\eta} . \label{56b}
\end{eqnarray}
\end{subequations}
Here, the functions $\kappa_1(\omega,q)$ and $\kappa_2(\omega,q)$ 
are the square roots of 
\begin{subequations} \label{57}
\begin{eqnarray}
\kappa_1^2 & = & q^2 + \frac{\omega_p^2}{c^2} - \frac{\omega^2}{c^2}
- i \frac{\omega_p^2}{c^2} \frac{\omega\eta}{\omega^2+\eta^2} ,\label{57a} \\
\kappa_2^2 & = & q^2 - \frac{\omega^2}{c^2} 
- i \frac{\omega^2}{c^2} {\rm sgn}(\omega) 0^+ \label{57b}
\end{eqnarray}
\end{subequations}
with a positive real part and the new function $\nu(\omega)$ is defined by
\begin{equation} \label{58}
\nu(\omega) = \omega^2 - 2 (cq)^2 \left\{
\begin{array}{cc}
<0 & \mbox{for $\omega^2=\omega_-^2$,} \cr & \cr
>0 & \mbox{for $\omega^2=\omega_+^2$.} \cr 
\end{array} \right.
\end{equation}
As indicated, only the sign of this function at the points 
$\omega^2=\omega_-^2$ and $\omega^2=\omega_+^2$ will be needed.

In order to keep the transparency of algebraic operations, we introduce
the counterparts of quantities $\beta S(\omega,q)$ (\ref{29a}) and 
$\beta S(t,q)$ (\ref{30}) for each of the components $g_1$ and $g_2$:
\begin{eqnarray} 
\beta S_j(\omega,q) & = & \frac{\beta\hbar}{4\pi} \coth(\beta\hbar\omega/2)
q^2 {\rm Im}\, g_j(\omega,q) , \label{59} \\
\beta S_j(t,q) & = & \int_{-\infty}^{\infty} \frac{d\omega}{2\pi}
e^{- i \omega t} \beta S_j(\omega,q) , \label{60}
\end{eqnarray}
$j=1,2$.
The quantity of interest is
\begin{equation} \label{61}
\beta S(t,q) = \beta S_1(t,q) + \beta S_2(t,q) .
\end{equation}
The evaluation of the introduced functions is the subject of the Appendix.
We summarize shortly the obtained results in the next paragraph. 

There are two linear in $q$ contributions to $\beta S_1(t,q)$.
The first one (\ref{A5}) originates from the discrete level of bulk
plasmons $\omega=\pm\omega_p$, the second one (\ref{A12}) results
from the integration over the continuous spectrum $\omega^2>\omega_p^2+(cq)^2$.
It is seen that the two contributions are exactly canceled with
one another, so that 
\begin{equation} \label{62}
\beta S_1(t,q) = o(q) .
\end{equation}
As concerns the quantity $\beta S_2(t,q)$, there is only one
linear in $q$ contribution (\ref{A20}) coming from the integration
over the continuous spectrum $\omega^2>(cq)^2$, i.e.
\begin{equation} \label{63}
\beta S_2(t,q) = \frac{q}{4\pi} + o(q) .
\end{equation}
We conclude that the total $\beta S(t,q)$ has the small-$q$ expansion
of the classical static type
\begin{equation} \label{64}
\beta S(t,q) = \frac{q}{4\pi} + o(q) .
\end{equation}
In other words, for both static and time-dependent surface charge
correlation functions, the inclusion of retardation effects causes 
the quantum prefactor to take its universal static classical form.
This result holds for any temperature.
The static $t=0$ version of (\ref{64}) is clearly consistent with 
the classical finding (\ref{6}), as it should be in the high-temperature 
limit $\beta\hbar\to 0$ \cite{Alastuey00}.
On the other hand, when $t\ne 0$, our result (\ref{64}) does not reproduce
the classical time-dependent one (\ref{13}), but this feature is not against 
the general principles discussed in the Introduction.

\section{Conclusion}
We have studied the long-range decay of the charge correlation function 
on the surface of the conductor in vacuum.
This problem has been investigated previously, for both static and 
time-dependent correlation functions, in the classical and quantum 
{\em nonretarded} regime.
Within the framework of the fluctuation EM-field theory we have shown that 
the consideration of retardation effects leads, for any temperature,
to the universal static classical form of the asymptotic decay:
\begin{equation} \label{65}
\beta S_{\rm qu}(t,\vert {\bf R}-{\bf R}'\vert) 
\mathop{\sim}_{\vert {\bf R}-{\bf R}'\vert\to\infty}
- \frac{1}{8\pi^2} \frac{1}{\vert {\bf R}-{\bf R}'\vert^3} ,
\end{equation}
independent of $t$ and $\hbar$.

As a model system for the conductor, we have used the jellium with
the simple dispersion relation of Drude type (\ref{34}).
It is not clear at the present stage whether the obtained result takes 
place also for other Coulomb or dielectric models with other types of
the dispersion relation.

As a byproduct of the formalism, we have obtained a very simple formula 
(\ref{41}) for the $g$-function, valid in the quantum nonretarded regime.
This formula reproduces the previous microscopic result for the jellium
and might be used to analyze general Coulomb or dielectric models,
characterized by their dielectric functions.
A general analysis might be possible also for the retarded case.

The extension of the present macroscopic study to other domain geometries,
e.g. a conductor confined to a slab, might be of interest. 

In the near future, we plan to perform a different analysis of a jellium 
coupled to the electromagnetic radiation, based on the method of 
collective modes developed in Refs. \cite{Jancovici85a,Jancovici85b}. 

\begin{acknowledgments}
L. \v{S}amaj is grateful to CNRS for supporting his stay at LPT.
A partial support by grant VEGA 2/6071/28 is acknowledged. 
\end{acknowledgments}

\appendix
*\section{}

\subsection{Contributions from $g_1$}
Let us first analyze the contributions of $g_1(\omega,q)$ 
to $\beta S_1(t,q)$, obtained by using Eqs. (\ref{59}) and (\ref{60}). 

Dividing $\kappa_1$ onto its real and imaginary parts, 
$\kappa_1=\kappa'_1+ i \kappa''_1$,
Eq. (\ref{57a}) splits into two relations
\begin{subequations} \label{A1}
\begin{eqnarray}
(\kappa'_1)^2 - (\kappa''_1)^2 & = & q^2 + \frac{\omega_p^2-\omega^2}{c^2} , 
\label{A1a} \\  \kappa'_1 \kappa''_1 & = & 
- \frac{\omega_p^2}{2 c^2} \frac{\omega\eta}{\omega^2+\eta^2} .
\label{A1b}
\end{eqnarray}
\end{subequations}
The function on the rhs of (\ref{A1b}) will help us to choose the correct 
sign of $\kappa''_1$ which is consistent with the condition $\kappa'_1>0$. 
We have to distinguish between two cases.
\begin{itemize}
\begin{subequations} \label{A2}
\item $\omega^2<\omega_p^2 + (cq)^2$:
\begin{equation} \label{A2a}
\kappa'_1 = \frac{1}{c} \sqrt{\omega_p^2+(cq)^2-\omega^2} , \qquad
\kappa''_1 = - \frac{1}{\kappa'_1} 
\frac{\omega_p^2}{2 c^2} \frac{\omega\eta}{\omega^2+\eta^2} ;
\end{equation}
\item $\omega^2>\omega_p^2 + (cq)^2$:
\begin{eqnarray} 
\kappa'_1 & = & - \frac{1}{\kappa''_1} 
\frac{\omega_p^2}{2 c^2} \frac{\omega\eta}{\omega^2+\eta^2} > 0 ,\nonumber \\ 
\kappa''_1 & = & - \frac{1}{c} {\rm sgn}(\omega)
\sqrt{\omega^2-\omega_p^2-(cq)^2} . \label{A2b}
\end{eqnarray}
\end{subequations}
\end{itemize}
Here and hereinafter, we adopt the convention that the square root 
of a real positive number has the plus sign and neglect all terms
of order $\eta^2$.
Since $\omega\eta/(\omega^2+\eta^2)\propto \omega\delta(\omega)$,
the integration over $\omega$ of the functions which contain this factor 
gives zero contribution. 

The function $g_1$, defined by Eq. (\ref{56a}), can be analyzed
by using the Weierstrass prescription (\ref{35}).
Since $\kappa_1(\omega_p)=q$ and 
$(\omega_p^2-\omega_-^2)(\omega_p^2-\omega_+^2) = - \omega_p^2 (cq)^2$,
we obtain
\begin{eqnarray}
{\rm Im}\, g_1 & = & - \frac{\pi\omega_p^2}{q} {\rm sgn}(\omega)
\delta(\omega^2-\omega_p^2) - (c\omega_p)^2 \omega^2 \kappa''_1 
\nonumber \\ & & 
\times {\cal P}\left( \frac{1}{\omega^2-\omega_p^2} \right)
{\cal P}\left( \frac{1}{\omega^2-\omega_-^2} \right)
{\cal P}\left( \frac{1}{\omega^2-\omega_+^2} \right) \nonumber \\ & &
+ \frac{\pi(c\omega_p)^2}{\omega^2-\omega_p^2} \omega^2 \kappa'_1\,
{\rm sgn}(\omega\nu)\delta[(\omega^2-\omega_-^2)(\omega^2-\omega_+^2)] .
\nonumber \\ & & \label{A3}
\end{eqnarray}
We take into account the lower bound (\ref{52}) for $\omega_+^2(q)$
and the upper bound (\ref{53}) for $\omega_-^2(q)$ and, as before, 
distinguish between two intervals.
\begin{itemize}
\begin{subequations} \label{A4}
\item $\omega^2<\omega_p^2 + (cq)^2$:
\begin{eqnarray} 
{\rm Im}\, g_1 & = & - {\rm sgn}(\omega) \frac{\pi\omega_p^2}{q} 
\delta(\omega^2-\omega_p^2) \nonumber \\ & &
- {\rm sgn}(\omega) \frac{\pi c\omega_p^2\omega_-^2}{(\omega_-^2-\omega_p^2)
(\omega_+^2-\omega_-^2)} \nonumber \\ & & \times
\sqrt{\omega_p^2+(cq)^2-\omega_-^2} \delta(\omega^2-\omega_-^2) ; \label{A4a}
\end{eqnarray}
\item $\omega^2>\omega_p^2 + (cq)^2$:
\begin{eqnarray} 
{\rm Im}\, g_1 & = & {\rm sgn}(\omega) 
\frac{c\omega_p^2\omega^2}{(\omega^2-\omega_p^2)(\omega^2-\omega_-^2)} 
\nonumber \\ & & \times \sqrt{\omega^2-\omega_p^2-(cq)^2}
{\cal P}\left( \frac{1}{\omega^2-\omega_+^2} \right) . \label{A4b}
\end{eqnarray}
\end{subequations}
\end{itemize}

The contributions of the discrete $\delta$-function terms in (\ref{A4a}) 
to $\beta S_1(t,q)$, defined by Eqs. (\ref{59}) and (\ref{60}),
are easy to find.
The bulk-plasmon term in (\ref{A4a}), which is proportional to 
$\delta(\omega^2-\omega_p^2)$, gives a contribution of order $q$:
\begin{equation} \label{A5}
- \frac{q}{4\pi} \cos(\omega_p t) f(\omega_p) ,
\end{equation}
where the function $f$ is defined by Eq. (\ref{8b}). 
The term in (\ref{A4a}), which is proportional to 
$\delta(\omega^2-\omega_-^2)$, gives a contribution of order $q^2$:
\begin{eqnarray}
- \frac{q^2}{4\pi} \cos(\omega_- t) f(\omega_-) 
\frac{c\omega_p^2}{(\omega_-^2-\omega_p^2)
(\omega_+^2-\omega_-^2)} \nonumber \\ \times 
\sqrt{\omega_p^2+(cq)^2-\omega_-^2}
= \frac{q^2}{4\pi} \frac{c}{\omega_p} + o(q^2) . \label{A6}
\end{eqnarray}
To find the contribution of the term in (\ref{A4b}), 
which is continuous in $\omega$, is a more complicated task.
In the small-$q$ limit, two functions $1/(\omega^2-\omega_p^2)$ and 
$\sqrt{\omega^2-\omega_p^2-(cq)^2}$ are close to their singular points 
just at the lower border of the integration $\omega^2=\omega_p^2+(cq)^2$.
Moreover, the principal value ${\cal P}(1/(\omega^2-\omega_+^2))$ 
has also to be taken with caution; see the small-$q$ expansion (\ref{54b}) 
of $\omega_+^2$. 
Based on the above information, we split the whole interval
of $\omega$-values $\omega^2>\omega_p^2+(cq)^2$ onto the small one 
symmetric with respect to $\omega_+^2$,
\begin{equation} \label{A7}
I_1:\ \omega_+^2-l < \omega^2 < \omega_+^2+l ,
\quad l = \omega_+^2 - [\omega_p^2+(cq)^2] \sim \frac{(cq)^4}{\omega_p^2} 
\end{equation} 
and the infinite one
\begin{equation} \label{A8}
I_2:\ \omega_+^2+l < \omega^2 .
\end{equation}
We first perform the integration of the three ``problematic''
quickly changing functions over the interval $I_1$.
Introducing the variable $u$ via $\omega^2=\omega_+^2+u$,
the contribution from the $\omega$-integration over the interval $I_1$ 
is the factor $(cq^2/4\pi^2)\cos(\omega_p t)(\beta\hbar\omega_p/2)
\coth(\beta\hbar\omega_p/2)$ multiplied by
\begin{eqnarray}
\int_{-l}^l d u \frac{1}{u+(cq)^2+l} \sqrt{u+l} \frac{u}{u^2+\eta^2} 
\nonumber \\
\sim \frac{1}{(cq)^2} \int_0^l d u \frac{1}{u} 
\left(\sqrt{l+u} - \sqrt{l-u} \right) \nonumber \\
= \frac{\sqrt{l}}{(cq)^2} 2 \left[ \sqrt{2} - \sinh^{-1}(1) \right] = O(1) .
\label{A9}
\end{eqnarray} 
This means that the contribution of (\ref{A4b}) to $\beta S_1(t,q)$ 
coming from the interval $I_1$ is of order $q^2$.
The integration of the term (\ref{A4b}) over the interval $I_2$
can be represented, with the substitution $\omega^2=\omega_+^2+l+v$
and in the small-$q$ limit, as follows 
\begin{eqnarray}
\frac{\beta\hbar q^2}{8\pi^2} c\omega_p^2 \int_0^{\infty} d v\,
\cos\left( t\sqrt{\omega_p^2+v}\right) \frac{\sqrt{v+2l}}{\sqrt{\omega_p^2+v}} 
\nonumber \\
\times \coth\left( \frac{\beta\hbar}{2}\sqrt{\omega_p^2+v}\right) 
\frac{1}{v+l} \frac{1}{v+(cq)^2} . \label{A10}
\end{eqnarray}
Performing the next substitution $v=(cq)^2 w$, considering the $q\to 0$ limit
and evaluating the integral
\begin{equation} \label{A11}
\int_0^{\infty} d w \frac{1}{\sqrt{w}} \frac{1}{w+1} = \pi ,
\end{equation}
the total contribution of the term (\ref{A4b}) to $\beta S_1(t,q)$ reads
\begin{equation} \label{A12}
\frac{q}{4\pi} \cos(\omega_p t) f(\omega_p) + o(q) .
\end{equation}
The linear in $q$ part of this contribution is exactly canceled
by the bulk-plasmon contribution (\ref{A5}).
We therefore conclude that
\begin{equation} \label{A13}
\beta S_1(t,q) = o(q) .
\end{equation}

\subsection{Contributions from $g_2$}
We proceed by a relatively simple analysis of the contributions 
of $g_2(\omega,q)$ to $\beta S_2(t,q)$. 

Writing $\kappa_2=\kappa'_2 + i \kappa''_2$, we have to distinguish between
two cases.
\begin{subequations} \label{A14}
\begin{itemize}
\item $\omega^2<(cq)^2$:
\begin{equation} \label{A14a}
\kappa'_2 = \frac{1}{c} \sqrt{(cq)^2-\omega^2} , \qquad \kappa''_2\sim 0 ;
\end{equation}
\item $\omega^2>(cq)^2$:
\begin{equation} \label{A14b}
\kappa'_2\sim 0^+ , \qquad
\kappa''_2 = -\frac{1}{c} {\rm sgn}(\omega) \sqrt{\omega^2-(cq)^2} .
\end{equation}
\end{itemize}
\end{subequations}

The function $g_2$ is defined by Eq. (\ref{56b}).
Using the Weierstrass theorem, we obtain
\begin{eqnarray}
{\rm Im}\, g_2 & = & (c\omega_p)^2 \kappa''_2
{\cal P}\left( \frac{1}{\omega^2-\omega_-^2} \right)
{\cal P}\left( \frac{1}{\omega^2-\omega_+^2} \right) \nonumber \\
& & - \pi (c\omega_p)^2 \kappa'_2 {\rm sgn}(\omega\nu)
\delta[(\omega^2-\omega_-^2)(\omega^2-\omega_+^2)] . \nonumber \\
& & \label{A15}
\end{eqnarray}
We thus have
\begin{subequations} \label{A16}
\begin{itemize}
\item $\omega^2<(cq)^2$:
\begin{equation} \label{A16a}
{\rm Im}\, g_2 = \pi c \omega_p^2 \sqrt{(cq)^2-\omega_-^2} {\rm sgn}(\omega)
\frac{1}{\omega_+^2-\omega_-^2} \delta(\omega^2-\omega_-^2) ;
\end{equation}
\item $\omega^2>(cq)^2$:
\begin{eqnarray} 
{\rm Im}\, g_2 & = & - c \omega_p^2 \sqrt{\omega^2-(cq)^2} {\rm sgn}(\omega)
\nonumber \\ & & \times
{\cal P}\left( \frac{1}{\omega^2-\omega_-^2} \right)
{\cal P}\left( \frac{1}{\omega^2-\omega_+^2} \right) . \label{A16b}
\end{eqnarray}
\end{itemize}
\end{subequations}

The contribution of the term (\ref{A16a}) to $\beta S_2(t,q)$ reads
\begin{eqnarray}
\frac{q^2}{4\pi} \cos(\omega_- t) f(\omega_-) 
\frac{c\omega_p^2}{\omega_-^2(\omega_+^2-\omega_-^2)} \nonumber \\ \times 
\sqrt{(cq)^2-\omega_-^2}
= \frac{q^2}{4\pi} \frac{c}{\omega_p} + O(q^4) . \label{A17}
\end{eqnarray} 
The contribution of the term (\ref{A16b}) can be represented,
after the substitution $u=\omega^2$ and in the small-$q$ limit, as follows
\begin{eqnarray}
\frac{\beta\hbar q^2}{8\pi^2} (-c\omega_p^2) 
\int_{(cq)^2}^{\infty} \frac{d u}{\sqrt{u}}
\cos(t\sqrt{u}) \coth(\beta\hbar\sqrt{u}/2) \nonumber \\ \times 
{\cal P}\left( \frac{1}{u-\omega_p^2} \right)
\frac{1}{\sqrt{u-(cq)^2}} . \label{A18}
\end{eqnarray}
Making the next substitution $u=(cq)^2 v$ and evaluating the integral
\begin{equation} \label{A19}
\int_1^{\infty} d v \frac{1}{v} \frac{1}{\sqrt{v-1}} = \pi ,
\end{equation}
the contribution of the term (\ref{A16b}) to $\beta S_2(t,q)$ is
found in the classical static form
\begin{equation} \label{A20}
\frac{q}{4\pi} + o(q) .
\end{equation}
In view of Eqs. (\ref{A17}) and (\ref{A20}) we conclude that
\begin{equation} \label{A21}
\beta S_2(t,q) = \frac{q}{4\pi} + o(q) .
\end{equation}

\bibliography{surface}

\end{document}